# Statistics of rupture in phantom chain network simulations

Yuichi Masubuchi*, Takato Ishida, and Takashi Uneyama

*Department of Materials Physics, Graduate School of Engineering, Nagoya University, Nagoya 464-8603, Japan*

*To whom correspondence should be addressed,

mas@mp.pse.nagoya-u.ac.jp

**Abstract**

Phantom chain simulations have shown that the mean rupture properties of star polymer networks collapse onto master curves against the cycle rank density $\xi$. This study revisits this universality with a much larger ensemble than in earlier studies to discuss the statistics. Phantom Gaussian networks were made by end-linking star prepolymers, and 1,000 realizations were collected for each of 30 conditions with functionality $f = 3 - 8$ and conversion $p = 0.60 - 0.95$, giving 30,000 networks in total. For each realization, the breaking stretch $\lambda_b$, the breaking stress $\sigma_b$, the breaking energy $W_b$, and the cycle rank $\xi$ were recorded. The master curves are unchanged by the larger sample, demonstrating that the earlier conclusions reported for the averages of smaller ensembles hold. However, the individual realizations are inherently random, and their statistical properties, rather than the individual values, are examined. At fixed $(f, p)$, the fluctuation of $\xi$ is small, varying by less than 1%, whereas $\lambda_b$, $\sigma_b$, and $W_b$ scatter by 5–30%. The fluctuation of $\xi$ is almost uncorrelated with that of the breaking properties. In addition, the scatter has a definite structure; its magnitude decreases with the mean cycle rank density $\langle\xi\rangle$, the $\lambda_b$–$\sigma_b$ correlation grows with $\langle\xi\rangle$, and the distributions deviate from Gaussian. The $\lambda_b$ distribution is skewed to the right at small $\langle\xi\rangle$, whereas $\sigma_b$ is skewed to the left at large $\langle\xi\rangle$. These rupture statistics were discussed in the framework of extreme-value statistics to demonstrate that the observed trends are opposite to those of the random fuse model, in which strength decreases with size and weakest-link statistics appear for weak disorder. The difference may reflect the source of fluctuation, i.e., the cross-linking in the present networks.

## 1. Introduction

Despite numerous studies, the structure-property relationship governing the rupture of cross-linked polymer networks has not yet been fully elucidated[1–4]. Among the simulation attempts [5], a recent series of phantom chain simulations [6–13] demonstrated that the cycle-rank density $\xi$, which is the number of independent loops per node [14–17], is a potential descriptor of network rupture. The studies were performed to end-link star prepolymers with various functionality $f$ and conversion $p$, and it was found that the breaking stretch $\lambda_b$, the breaking stress $\sigma_b$, and the breaking energy $W_b$ collapsed onto master curves when plotted against $\xi$. The $\lambda_b - \xi$ relation has been explained by a mechanical model [18], in which the rupture path is replaced by a soft bottleneck strand in series with stiffer parallel bundles, and the parallelism is related to $\xi$.

The finding about the role of $\xi$ raises a few questions. The first natural one is whether $\xi$ can predict rupture in solo. Note that the above $\xi$-universalities have been reported for the average values among several simulation runs for the given sets of $f$ and $p$, and not for individual networks. Indeed, the recent simulation study by Ishikura et al.[19] on various randomly cross-linked bead-spring networks addresses this issue and suggests that additional parameters, such as the effective chain ratio and strand-length uniformity, are necessary. The second question is about the rupture statistics. The statistical description of fracture in disordered media has a long history[20]; when failure is governed by the weakest of elements, the breaking property is an extreme value of many random variables. The Fisher–Tippett–Gnedenko theorem states that the distribution of such an extreme value converges to one of three limiting forms, namely the Gumbel, Fréchet, or Weibull distribution, irrespective of the details of the elements.[21] In this framework, the standard models are lattice models, such as the random fuse model (RFM) and the fiber-bundle model, where a network of elements with random breaking thresholds is considered. In contrast, in simulations of polymer networks, all strands are identical, and the source of realization-to-realization variation is the structural fluctuation generated by the random cross-linking reaction. Since this intrinsic fluctuation differs from the externally imposed threshold disorder in the RFM, the rupture statistics may also differ. However, no literature has addressed this issue.

In this study, a large ensemble is generated, with 1000 realizations for each of 30 conditions at various $f$ and $p$. The within-condition scatter of the breaking properties is examined, together with the question of whether $\xi$ predicts that scatter, and the shape of the distributions. First, it is confirmed that the larger sample does not change the master curves. Namely, the earlier average-level conclusions are not artifacts of small samples. Then the structure of the scatter is reported, and is compared with the RFM.

## 2. Simulations

The simulation protocol follows previous studies[6–10,12,13,22], and a brief description is provided here. Star prepolymers with $f$ arms of $N_a = 5$ beads each were dispersed in a periodic simulation box at a fixed bead number density $\rho$. The polymers are Rouse–Ham-type and the beads are connected by Gaussian springs. End-linking reactions were turned on after equilibration, and primary loops were eliminated by disallowing reactions within the same prepolymer. During the gelation simulations, network snapshots were stored at several conversions $p$. The stored networks were uniaxially stretched via sequential energy minimization and infinitesimal step strain under volume conservation[12]. In each energy-minimized structure, the bond length was examined, and the bond was removed when its length exceeded a critical length $b_c$. The breaking stretch $\lambda_b$ was taken from the peak of the nominal stress–strain curve, the breaking stress $\sigma_b$ from that peak, and the breaking energy $W_b$ from the area under the curve up to $\lambda_b$. The cycle rank $\xi$ was computed for the percolating cluster.

The functionality was $f$ = 3, 4, 5, 6, 7, 8, and the conversion was $p$ = 0.60, 0.70, 0.80, 0.90, 0.95, giving a 6×5 grid of 30 conditions. Units of length and energy are the equilibrium bond length before energy minimization, and thermal energy. The simulation parameters were chosen at $\rho$ = 8 and ${b_c}^2$=1.5. The system included ca. 33,000 beads, which slightly varied according to f. For each condition, 1000 realizations were performed, totaling 30000 networks. The ensemble is more than two orders of magnitude larger than those in previous studies, which used eight realizations per condition, and it is necessary to examine the within-condition distributions. Hereafter, $\langle \cdot \rangle$ denotes the average over the 1000 realizations for each $(f, p)$ condition. As a stress scale, the branch-point number density $\nu_{br} = \rho/(fN_a + 1)$ is used, and $\sigma_b$ and $W_b$ are normalized by it where needed.

## 3. Results and Discussion

Figure 1 shows every realization plotted against its cycle rank, together with the conditional means, for $\lambda_b$, $\sigma_b/\nu_{br}$, and $W_b/\nu_{br}$. The means reproduce the previously reported master curves [5,6,10,13)]. Namely, $\lambda_b$ decreases with $\xi$, while $\sigma_b/\nu_{br}$ and $W_b/\nu_{br}$ increase, and the six functionalities fall on common trends. With 1000 realizations, the means are essentially noise-free. Thus, the earlier conclusions for the averages, drawn from much smaller samples, are confirmed.

Meanwhile, the individual realizations show large fluctuations around the master curves. For each condition, the fluctuation of $\xi$ is small at a given $(f, p)$, while the breaking properties spread around the mean, with the largest spread for $W_b$. Since the individual realizations are inherently random, their statistical properties, rather than the individual values, must be characterized. In the following, the fluctuations and the distribution functions of the breaking properties are examined.

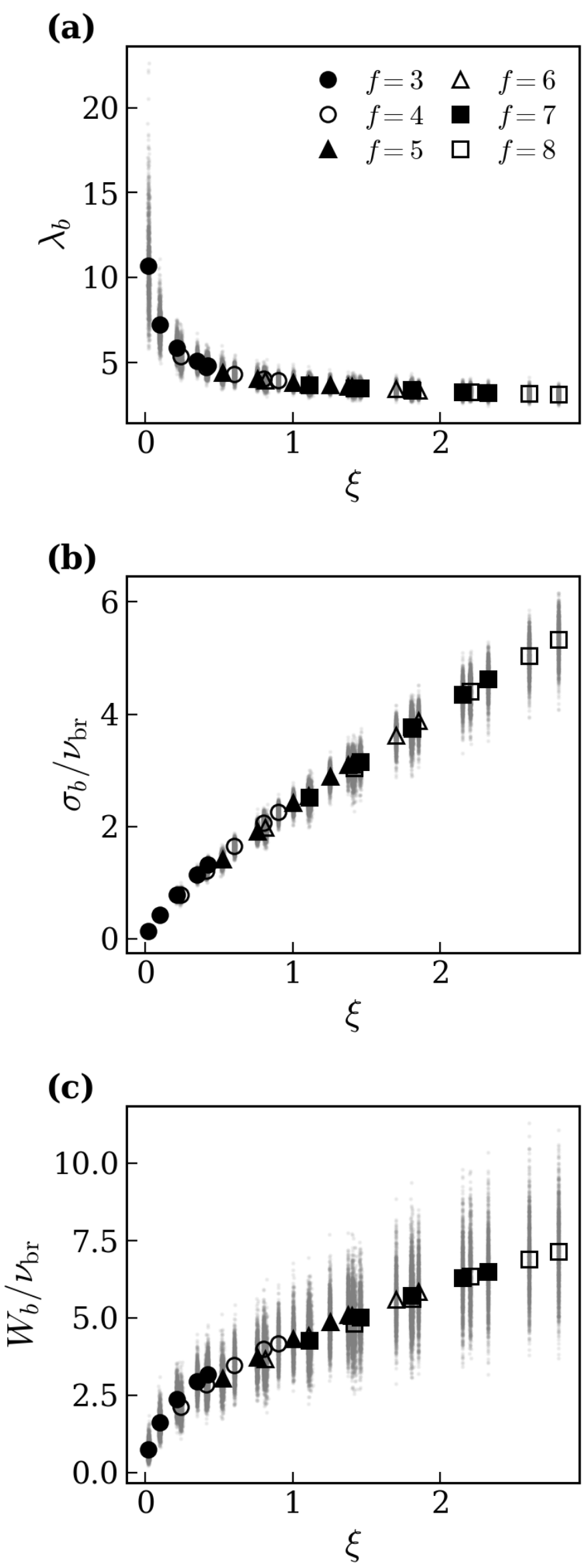

**Figure 1.** Individual realizations (gray points) and conditional means (symbols) against the cycle rank $\xi$ for $f$ = 3 (filled circle), 4 (open circle), 5 (filled triangle), 6 (open triangle), 7 (filled square), and 8 (open square). (a) breaking stretch $\lambda_b$, $\lambda_b$ (b) breaking stress normalized by the branch-point density $\sigma_b/\nu_{br}$, and (c) breaking energy $W_b/\nu_{br}$.

The fluctuation of $\xi$ in Fig. 1 is quantified in Fig. 2(a) by the within-condition relative standard deviation RSD($\xi$). The value is below 1% for all but the sparsest networks with $f = 3$, and below 0.05% at high $p$. This small fluctuation reflects that $\xi$ is an average over the network structure, whose relative fluctuation decreases as the system size grows, following the central limit theorem. Under the present conditions, the system is large enough to keep the fluctuation of $\xi$ below 1%. Therefore, the fluctuation of $\xi$ cannot account for that of the breaking properties, whose RSD is 6–23% for $\lambda_b$, 4–21% for $\sigma_b$, and 13–33% for $W_b$. Figure 2(b) shows the squared correlation coefficient between $\xi$ and $\lambda_b$ or $\sigma_b$ within the same $(f, p)$. Over the examined 30 conditions, the median is below 0.01, clearly demonstrating that the fluctuation of $\xi$ does not account for that of the breaking properties. The maximum value is 0.22 at $f$ = 3, $p$ = 0.6, near percolation.

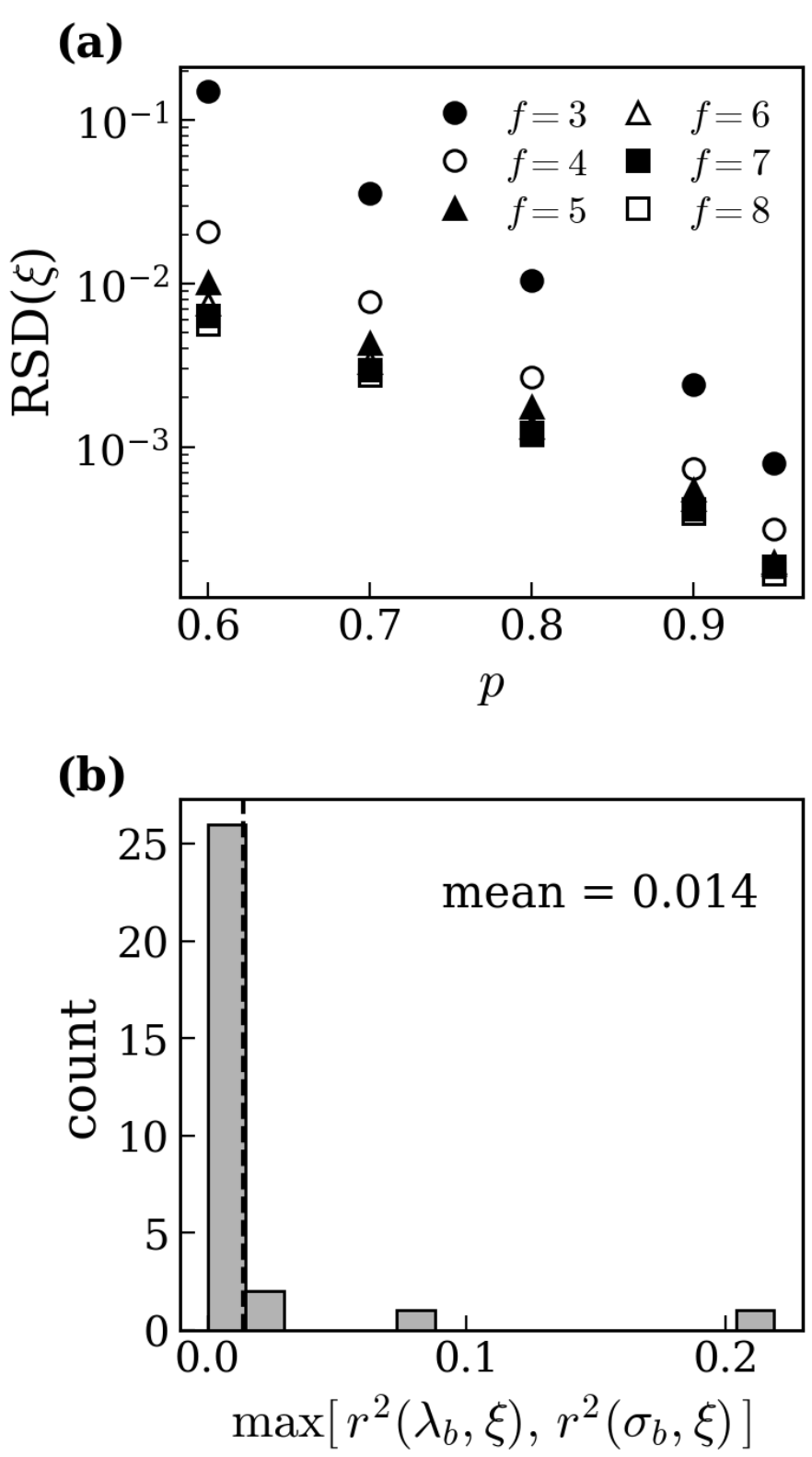

**Figure 2.** (a) The within-condition relative standard deviation RSD($\xi$) against the conversion $p$. Symbols as in Fig. 1. (b) Histogram, over the 30 conditions, of the squared correlation coefficient between $\xi$ and $\lambda_b$ or $\sigma_b$. The broken line indicates the mean, 0.014.

Although $\xi$ does not account for the fluctuation of individual networks, the statistics of the fluctuation depend on the mean cycle rank $\langle \xi \rangle$. Figure 3 shows the within-condition Pearson correlation coefficient between $\lambda_b$ and $\sigma_b$, computed over the 1000 realizations for the same $(f, p)$. It rises from near zero at small $\langle \xi \rangle$ to about 0.7 at large $\langle \xi \rangle$, and the data for different $f$ fall on a common trend. This result demonstrates that the two breaking properties fluctuate almost independently at low connectivity, while at high connectivity, a stronger realization is also more extensible.

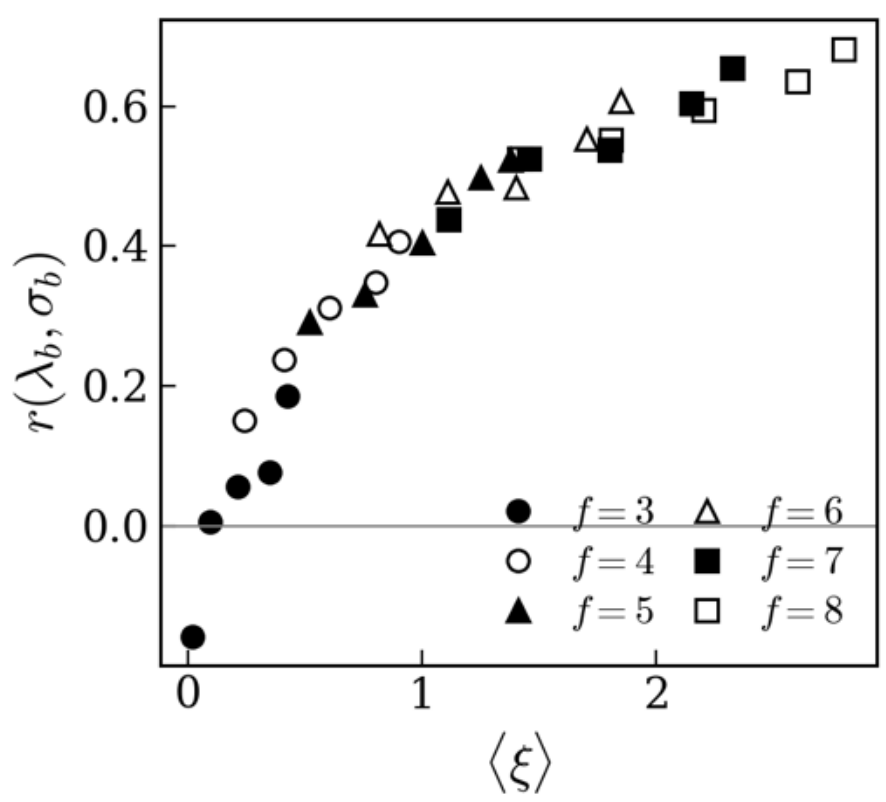

**Figure 3.** The within-condition correlation $r(\lambda_b, \sigma_b)$ against the mean cycle rank $\langle \xi \rangle$ for $f$ = 3 (filled circle), 4 (open circle), 5 (filled triangle), 6 (open triangle), 7 (filled square), and 8 (open square).

Figure 4 shows the distributions for three representative conditions over the examined range of $\langle \xi \rangle$. Since the breaking properties are extreme values, extreme-value distributions are natural candidates for describing the data[21)]. However, the shape parameter could not be reliably determined from the 1,000 realizations because it fluctuated strongly during resampling, and the ensemble is too small to identify the extreme-value family, whose tails must be resolved. Therefore, the distributions were fitted to the Gaussian, log-normal, and Weibull (minimum-value) forms, and the best form was selected by the Akaike information criterion. The stability of the selection was examined using bootstrap resampling, and normality was assessed separately with the Shapiro–Wilk test. These tests show that the distributions are not simply Gaussian and that their shapes change with $\langle \xi \rangle$. For $\lambda_b$ (top row), the distribution is skewed to the right at small $\langle \xi \rangle$, and a Gaussian cannot follow the long right tail. It becomes more symmetric at large $\langle \xi \rangle$. For $\sigma_b$ (middle row), the distribution is nearly symmetric at small $\langle \xi \rangle$, and it develops a left skew with a sharp upper edge at large $\langle \xi \rangle$, where the minimum-value Weibull form realizes the best fit. This shape is reminiscent of weakest-link statistics, in which failure is set by the weakest of many parallel elements. In contrast, $\Sigma = \sigma_b/\lambda_b$ (bottom row) is close to Gaussian throughout. The skewnesses of $\lambda_b$and $\sigma_b$, which are opposite in sign, appear to partly cancel in the ratio.

Note that the observation above depends on the sample size because resampling may alternate the best-fitting functional form among the candidates when the skewness is comparable to its standard error. Further calculations with a larger ensemble are required for such cases to identify the behavior of the tails. In contrast, the strongly skewed cases are robust; the skewness of $\lambda_b$ at small $\langle \xi \rangle$ and the negative skewness of $\sigma_b$ at large $\langle \xi \rangle$ are sufficiently larger than the standard errors, and these deviations from Gaussian are statistically secure, and the best-fitting family is stable under resampling.

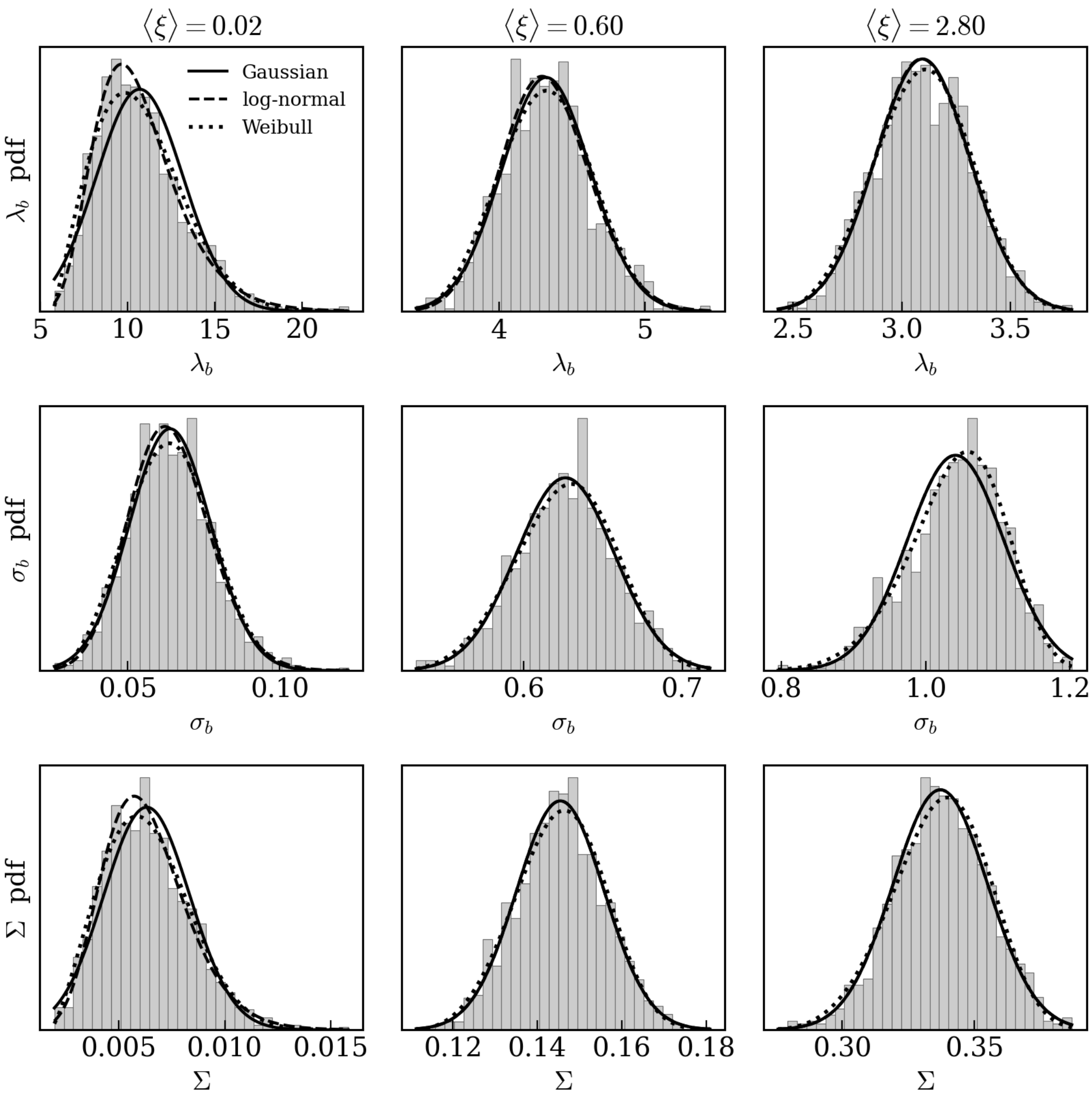

**Figure 4.** Distributions of $\lambda_b$ (top), $\sigma_b$ (middle), and $\Sigma = \sigma_b/\lambda_b$ (bottom) for three representative conditions of mean cycle rank (left to right: $\langle\xi\rangle$ = 0.02, 0.60, 2.80). Gray histograms are the data with 1,000 realizations each. Curves are the Gaussian (solid), log-normal (broken), and Weibull minimum-value (dotted) fits.

The observation on the distribution functions is summarized by the skewness shown in Fig. 5(a). The skewness of $\lambda_b$ is positive at small $\langle\xi\rangle$ and decreases toward zero. Meanwhile, the skewness of $\sigma_b$ starts near zero and becomes increasingly negative. The two move in opposite directions, indicating that the two breaking properties are shaped by different parts of the fluctuation. Namely, $\lambda_b$ is skewed to the right by a few highly extensible realizations, and $\sigma_b$ is skewed to the left by the upper bound that appears as the number of parallel elements increases.

Figure 5(b) shows the relative standard deviation of $\Sigma = \sigma_b/\lambda_b$. As shown in Fig. 4, $\Sigma$ is close to Gaussian in most conditions, in contrast to the skewed distributions of $\lambda_b$ and $\sigma_b$. Its fluctuation is therefore well characterized by the RSD, which decreases with $\langle\xi\rangle$, and the behavior differs between the low and high $\langle\xi\rangle$ regions, as seen in the difference in the apparent slopes. Since the breaking properties are extreme values of the network response, their fluctuations are expected to follow extreme-value statistics rather than the central limit theorem, and a single power law is not appropriate. The two regimes and the origin of this $\langle\xi\rangle$-dependence are left for future study.

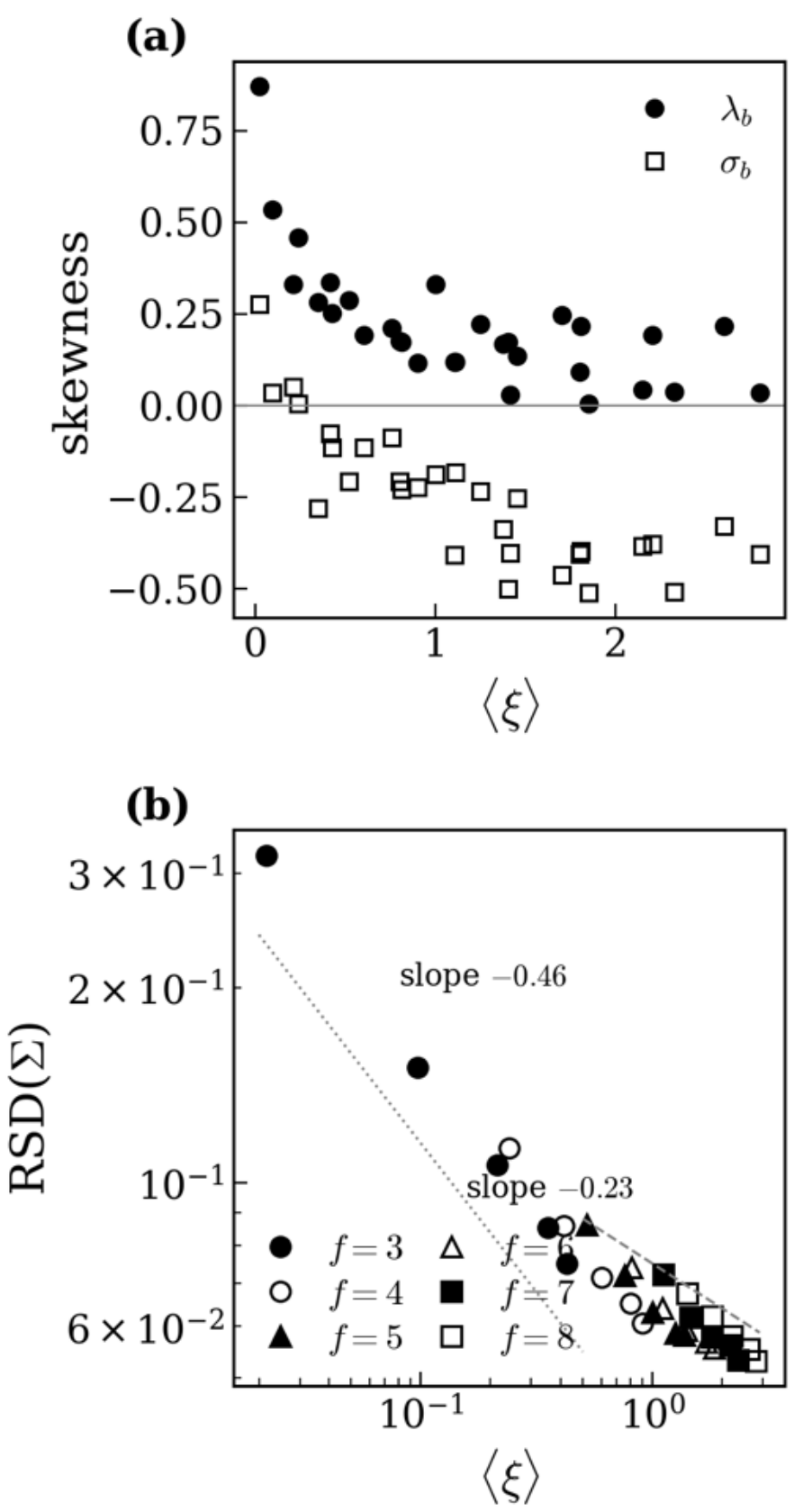

**Figure 5** (a) Skewness of the $\lambda_b$ (filled circles) and $\sigma_b$ (open squares) distributions and (b) the relative standard deviation of $\Sigma = \sigma_b/\lambda_b$ for $f$ = 3 (filled circle), 4 (open circle), 5 (filled triangle), 6 (open triangle), 7 (filled square), and 8 (open square) against $\langle\xi\rangle$. The dotted and broken lines in panel (b) indicate apparent slopes of $-0.46$ and $-0.23$ as visual guides, not fits.

The results reveal differences in rupture statistics relative to RFM, reflecting the fundamental distinction between the externally imposed threshold disorder and the intrinsic structural fluctuations of the present networks. For RFM, the shape of the strength distribution depends on the disorder; broad disorder yields a log-normal distribution, arising from the multiplicative accumulation of damage up to the peak load[23,24]. In the present fluctuation-driven case, the breaking-stress distribution is log-normal at low $\xi$, and a minimum-value form appears only at high $\xi$, where a considerable number of parallel paths exist; more connectivity brings on weakest-link behavior. The magnitude of the fluctuation also behaves differently. In the RFM, failure at the brittle limit is governed by a single critical defect, and strength fluctuations follow the Duxbury–Leath–Beale statistics specific to the diluted limit[25]. In the present study, RSD decreases as $\langle\xi\rangle$ grows, as shown in Fig. 5(b), implying that the failure is shared among multiple paths. In contrast, one apparent coincidence is that a log-normal breaking-stress distribution appears at low $\xi$ in this study and also in the RFM for broad disorder. However, their origins are different. In the RFM, it arises from the multiplicative accumulation of damage, whereas here it arises from a few dangling-rich realizations. Whether these observations reflect a real difference in universality or are only due to the limited range of this study would require assessing size dependence and the distribution tails. For instance, in the RFM, the mean strength decreases with system size $L$ because a larger system is more likely to contain a critical defect [26,27]. In the diluted limit, the strength falls slowly with $L$, and the distribution follows the Duxbury–Leath–Beale form[26,27], which is not one of the three limiting forms but flows to the Gumbel form for large systems[25]. These characteristics must be carefully tested in the present model with a larger study for further discussion.

## 6. Concluding remarks

An ensemble of 30,000 phantom network fracture simulations was used to investigate the statistics of the rupture properties. The mean values collapse onto master curves as reported for smaller ensembles, showing the robustness of the reported universalities. However, the individual realizations are inherently random, and their fluctuations were characterized instead of the individual values. The fluctuation of $\xi$ at fixed $(f, p)$ is small, whereas it is almost uncorrelated with the breaking properties and explains less than 1% of their within-condition variance. Regarding rupture statistics, as the mean cycle rank increases, the fluctuation narrows, the breaking stretch and stress become more correlated, and the distributions deviate from Gaussian. Specifically, the distribution for $\lambda_b$ is skewed to the right at low $\langle\xi\rangle$, and that for $\sigma_b$ is skewed to the left at high $\langle\xi\rangle$. Since the breaking properties are extreme values of the network response, extreme-value statistics provide a natural framework; however, the present ensemble is too small to identify the extreme-value family, whose tails must be resolved. Consequently, the present study distinguishes two roles of the cycle rank: it predicts the mean values of rupture properties and organizes their fluctuation as a mean-field coordinate. However, it does not fix the outcome of any single network. This finding is consistent with the view that the cycle rank alone is insufficient and that a second descriptor is needed to characterize the fracture of individual networks[19,28–30], even for the examined star-polymer networks with monodisperse strand length.

Concerning the statistics, qualitative differences were observed from those established for RFM[23–25]; the magnitude of fluctuation decreases with $\langle\xi\rangle$, the $\lambda_b$–$\sigma_b$ correlation grows with $\langle\xi\rangle$, and the $\lambda_b$ distribution is skewed to the right at small $\langle\xi\rangle$, whereas $\sigma_b$ is skewed to the left at large $\langle\xi\rangle$. Although these results hint at differences between threshold and fluctuation-driven networks, further investigations are required with much larger ensembles. Supplementary studies are ongoing, and the results will be published elsewhere.

## Acknowledgements

This study was partly supported by the Eno Science Foundation and JSPS KAKENHI (26H02291).